\documentclass[12pt,preprint]{aastex}

\shorttitle{Diffuse X-ray Background toward MBM20 with {\it Suzaku}}
\shortauthors{A. Gupta, M. Galeazzi, R. Smith, D. Koutroumpa, and R. Lallement}

\begin{document}
\title{Properties of the Diffuse X-ray Background toward MBM20 with {\it Suzaku}}


\author{A. Gupta and M. Galeazzi}
\affil{Physics Department, University of Miami, Coral Gables, FL 33124}
\email{galeazzi@physiscs.miami.edu}
\author{D. Koutroumpa}
\affil{NASA/GSFC, Code 662, Greenbelt, MD 20771}
\author{R. Smith}
\affil{Harvard Smithsonian Center for Astrophysics, Cambridge, MA 02138}
\author{R. Lallement}
\affil{CNRS Service d'Aéronomie, FR 91371 Verrieres Buisson, France}


\begin{abstract}

We used {\it Suzaku} observations of the molecular cloud MBM20 and a low 
neutral hydrogen column density region nearby to separate and 
characterize the foreground and background diffuse X-ray emission. 
A comparison with a previous observation of the same regions with 
{\it XMM-Newton} indicates a significant change in the foreground flux 
which is attributed to Solar Wind Charge eXchange (SWCX). 
The data have also been compared with previous results from 
similar ``shadow'' experiments and with a SWCX model to characterize 
its O~{\tiny VII} and O~{\tiny VIII} emission.
\end{abstract}



\section{Introduction}

Our current interpretation of the diffuse X-ray emission below 1 keV 
includes a combination of 5 components: Solar Wind Charge eXchange (SWCX), 
Local Bubble (LB), Galactic Halo (GH), Warm-Hot Intergalactic 
Medium (WHIM), and unresolved point sources (e.g., \citep{Gupta09},
\citep{Galeazzi09}). 
Resolving the different components is made particularly difficult due to 
the similar spectra of the components, primarily X-ray lines from heavily 
ionized metals. Compounding the problem, these lines are poorly resolved 
by the CCD cameras onboard current X-ray satellites.

Shadowing observations offer a tool to simplify the study of the various 
components by separating local (less than about 100 pc) and non-local 
components. A typical shadow experiment consists of two observations, 
one in the direction of a high latitude, high neutral hydrogen density 
cloud at a distance of 50 - 200 pc, the other toward a low neutral 
hydrogen column density sightline as close as possible to the cloud. 
As the cloud absorbs most of the background X-ray emission, a 
comparison of the two observations separates the 
foreground (LB plus SWCX) and background (GH, WHIM, and unresolved 
point sources) emission. Simultaneous spectral analysis of the two 
observations then determines the physical parameters of 
the different components. 

The results, however, are complicated by the properties of the SWCX 
component, which varies both in spectra composition and flux on 
a scale of hours to days. The SWCX component originates in the 
interaction of the highly ionized solar wind with neutral gas in 
the Earth's atmosphere and in the interplanetary medium. 
The charge exchange occurs when an electron jumps from the neutral 
atom to an excited level in the ion. The electron then cascades to 
the lower energy level of the ion, emitting soft X-rays and other 
lines in the process. 
While the properties of the SWCX emission are not fully 
understood, it seems quite clear that, for a proper comparison of 
on-cloud and off-cloud data, the two observations must be temporally 
as close as possible. Even in such conditions, however, a careful 
monitoring and simulation of the expected SWCX emission is needed.

We have obtained spectra of the Soft X-Ray Background (SXRB) in the 
direction of the high latitude, neutral hydrogen cloud MBM20 and a 
low neutral hydrogen column density region nearby that we called the 
Eridanus Hole (EH) using the X-ray Imaging Spectrometer 
(XIS; \citep{Koyama07}) 
onboard the {\it Suzaku} X-ray observatory. The XIS is an excellent tool 
for studying the SXRB, due to its low and stable non-X-ray background 
and good spectral resolution. The targets are identical to those 
observed in Galeazzi et al. (2007) using {\it XMM-Newton}.

To build a consistent picture of the diffuse X-ray background we 
compared the {\it Suzaku} observation with the previous {\it XMM-Newton} result 
and with similar shadow observations in the direction of the high 
latitude molecular cloud MBM12 \citep{Smith05, Smith06} and a filament 
in the southern Galactic hemisphere \citep{Henley07, Henley08}. We also 
used the model recently developed by Koutrompa et al. (2007) to estimate 
the emission from SWCX. The model is time dependent and includes factors 
such as solar cycle phase, the observation position, and the line of sight.

The data reduction is discussed in \S 2 and the analysis and X-ray 
results in \S 3. Section 4 compares our result with the previous 
{\it XMM-Newton} observation generally with other recent shadow experiments, 
and \S 5 to the characterization of SWCX, including models of the 
SWCX emission during the {\it Suzaku} and {\it XMM-Newton} observations.

\section{Observations of MBM20 and Eridanus Hole}

MBM20 and the Eridanus Hole were observed with {\it Suzaku} in February 2008 
and July 2007 respectively. Notice that the temporal gap between the 
observations is large compared to the typical time variation of the 
SWCX and will be discussed in section 5.The details of the observations 
are reported in Table 1. MBM20 is a high-density, high-latitude 
star-forming cloud located at or within the edge of the Local Bubble
\citep{Galeazzi07}. 
Its mass is ~84M$_\odot$ and it is located at coordinates 
$\it {l}=211^\circ23'53''.2$, $\it {b}=-36^\circ32'41''.8$, southwest 
of the Orion star forming complex. Based on interstellar NaI D 
absorption lines the distance of MBM20 is evaluated between 
$112\pm15$~pc and $161\pm21$~pc \citep{Hearty00}. The Eridanus hole, 
at coordinates  $\it {l}=213^\circ25'52''.3$, 
$\it {b}=-39^\circ5'26''.6$, is a region of low neutral hydrogen 
column density located about 2 degrees from the highest-density part 
of MBM20 (Fig.~1).

\subsection{Data Reduction}

We used the {\it Suzaku} data reprocessed to version 2.0 and the analysis 
was performed with HEAsoft\footnote{See http://heasarc.nasa.gov/lheasoft/} 
version 6.4 and XSPEC 12.4.0. We started the event screening from the 
cleaned event file, in which selection of the event grade and bad CCD 
column, and removal of hot and flickering pixels by the ``cleansis'' 
ftool, were already conducted ({\it Suzaku} Data Reduction 
Guide\footnote{See http://suzaku.gsfc.nasa.gov/docs/suzaku/analysis/abc/abc.html/}).

In our analysis, we use only data from the XIS1 detector, as this 
has the greatest sensitivity at low energies. We combined the data taken in 
the 33 and 55 observation mode. For that, first we convert the $5\times5$ 
mode data to $3\times3$ mode data using Ftool ``xis$5\times5$to$3\times3$'', 
then merged both files with the help of Ftool ``ftmerge''.  The cleaned 
event files are by default filtered to exclude times within 436 seconds of 
{\it Suzaku} passing through the South Atlantic Anomaly (SAA), and when {\it Suzaku}'s 
line of sight is elevated above the Earths limb by less than $5^\circ$, or 
is less than $20^\circ$ from the bright-Earth terminator. We decided to 
expand this to exclude events with Earth-limb elevation angle less than 
$10^\circ$, as there are some excess events in the 0.5-0.6 keV band in 
the $5^\circ-10^\circ$ range \citep{Smith06}.

Due to {\it Suzaku}'s broad point spread function (half-power diameter 
$\sim$2'; \citep{Mitsuda07}), it is hard to detect point sources. 
Therefore, to remove sources which could 
contaminate our SXRB spectra, we used the location of sources determine 
in {\it XMM-Newton} observations (see Fig.~2). We extracted spectra from the 
full XIS1 field of view, after removing above-mentioned point sources and 
the corners of the detectors which contained the onboard Fe-55 calibration 
sources. 

\subsection{Background Removal}

{\it Suzaku} is in a low-Earth orbit, so it is significantly shielded from 
the particle background that strongly affects {\it XMM-Newton} and {\it Chandra}.  
The effectiveness of this shielding is dependent upon the 
``cut-off rigidity'' (COR) of the Earth's magnetic field, which varies 
as {\it Suzaku} traverses its orbit. During times with larger COR values, 
fewer particles are able to penetrate to the satellite and to the XIS 
detectors. We excluded times when the COR was less than 8 GV, which is 
higher than the default value (COR$\>$4 GV) for both observations, as the
lowest background was desired.

Although it is reduced by the Earth's magnetic field, {\it Suzaku} still has 
a noticeable particle background. We can estimate the appropriate 
particle background from a database of the night Earth data (NXB). 
NXB was collected when the telescope was pointed at the night Earth 
(elevation less than -5 degree, and pointed at night side rather than day). 
The event files in the database have been carefully screened for telemetry 
saturation and other artifacts. 
We constructed the spectra of the night earth data 
using Ftool ``xisnxbgen'' \citep{Tawa08}, which sorts the NXB data by COR 
values, generates an NXB spectrum and image for each COR range, and 
combines them weighted by exposure time ratio of each COR range during 
GTIs in our spectral data file. The background spectra were then subtracted 
from the corresponding source spectra. 

\subsection{XIS Response}

We calculated the XIS detector effective area using the tool 
``xissimarfgen'' \citep{Ishisaki07}. This tool takes into account the 
spatially varying contamination on the optical blocking filters of the 
XIS sensors which reduces the detector efficiency at low energies 
\citep{Koyama07}. For the ancillary response file (ARF) calculations 
we assumed a uniform source of radius 20' and used a detector mask which 
removed the bad pixel regions. To generate the redistribution matrix 
file (RMF), we used the ftool ``xisrmfgen''.

\section{Analysis}

We first fit a model to our spectra consisting of 3 components: 
a Local Bubble component, modeled as an unabsorbed plasma with 
thermal emission in collisional ionization equilibrium (CIE); 
a hotter Galactic halo emission, modeled as an equilibrium thermal 
plasma component absorbed by the gas in the Galactic disk; and an 
unresolved extragalactic source component, modeled with an absorbed 
power law. This is the same model used in Galeazzi et al. (2007).
As extensively discussed in \S\S4\&5, the data are affected
by ``contamination'' due to SWCX which limits the significance
of the results obtained with this model. However, with the limited
energy resolution of the CCD detectors the mentioned model works
quite well and allows for a straightforward comparison with previous
results. 

We used the XSPEC v12.4 (Arnaud \& Dormer 2002) to fit both spectra, 
in the energy range 0.4-5.5 keV. For plasma thermal emission, the 
Astrophysical Plasma Emission Code (APEC) was used \citep{Smith01}, 
and for the absorption, we used the XSPEC {\it wabs} model, which uses 
cross-sections from Wisconsin (Morrison \& McCammon, 1983) and uses 
the Anders \& Ebihara (1982) relative abundances. We fit the above 
mentioned model to the {\it Suzaku} spectra of MBM20 and the Eridanus Hole. 
As in Galeazzi et al. (2007), we used the $IRAS$ 100$\mu$m maps to 
evaluate the neutral hydrogen density in the two regions. The $IRAS$ 
average brightness is 13.34~MJy~sr$^{-1}$, and 0.73~MJy~sr$^{-1}$ for 
MBM20, and the Eridanus hole respectively. Using the ``typical'' 
high-latitude 100 $\mu$m/NH ratio of 
$0.85\times10^{-20}$~cm$^2$~MJy~sr$^{-1}$ (Boulanger \& Perault 1988) 
the estimated neutral hydrogen densities are 
$1.59\times10^{21}$cm$^{-2}$, and 0.86$\times10^{20}$cm$^{-2}$ respectively. 
The fits are shown in Fig.~3, along with the best-fitting multicomponent 
spectral model. The model parameters are reported in Table~2.

We also tried to fit the above mention model simultaneously to our MBM20 
and Eridanus Hole {\it Suzaku} spectra with a single set of parameters, except 
for the neutral hydrogen column density. The fits are shown in Fig.~4, 
and the model parameters are presented in Table~2.

To extend the analysis further, we also included data from the {\it ROSAT} 
ALL-Sky Survey (RASS) in the same directions. We extracted RASS data 
in the {\it ROSAT} bands R1-R7 \citep{Snowden98} and scaled them to the same 
field of view as our {\it Suzaku} data sets for both MBM20 and the Eridanus Hole. 
We then performed a global fit of the four data sets simultaneously with 
a single set of parameters. The fit results are reported in Table~2, 
and the data are shown in Fig.~5. Overall, the model gives a good fit 
to the data (reduced $\chi^{2}=0.87$ for 352 degrees of freedom), however, 
the fit to some of the {\it ROSAT} bands is rather poor.  

We used our fit results to obtain O~{\tiny VII} and O~{\tiny VIII} 
intensities, since at temperatures of million Kelvins, O~{\tiny VII} 
and O~{\tiny VIII} lines are the dominant features. 
In our {\it Suzaku} spectra of MBM20 and Eridanus Hole, the blended O~{\tiny VII} 
triplet at 561, 569 and 574~eV is clearly visible in both observations, 
while the O~{\tiny VIII} line at 654~eV is barely visible in the MBM20 
data set and lies within the statistical uncertainty in the Eridanus Hole 
data set. The O~{\tiny VII} and O~{\tiny VIII} line intensities are 
$2.26\pm0.6$ 
$\textrm{photons}~\textrm{s}^{-1}~\textrm{cm}^{-2}~\textrm{sr}^{-1}$ 
(line units, LU, from now on) and $0.56\pm0.48$ LU for MBM20, 
and $5.68\pm1.04$ LU and $1.32\pm0.79$ LU for the Eridanus hole respectively. 
Following the same recipe used in Galeazzi et al. (2007), we can 
evaluate O~{\tiny VII} and O~{\tiny VIII} emission of the foreground 
(LB+SWCX) and background (GH) components. Using the expression for cross 
section per hydrogen atom for a cosmic abundance plasma derived by 
Morrison \& McCammon (1983), we find 
that MBM20 absorbs about 75\% of the background O~{\tiny VII} emission 
and about 61\% of the background O~{\tiny VIII} emission, while the 
Eridanus Hole absorbs about 8\% of the background O~{\tiny VII} emission 
and about about 5\% of the background O~{\tiny VIII} emission. 
Combining these data with the result of our observations we obtain, 
for O~{\tiny VII} and O~{\tiny VIII} respectively, $0.99\pm0.91$ LU and 
$0.014\pm1.01$ LU for the foreground and $5.10\pm1.79$ LU and 
$1.42\pm1.74$ LU for the background.

We evaluated the electron density and thermal pressure of the GH and 
the LB, using the same procedure discussed in Galeazzi et al. (2007). 
Assuming the foreground component is due solely to LB emission, we 
obtain lower and upper limits for the plasma density of 
0.015 and 0.018 $cm^{-3}K$ and limits of 23,500 and 28,800 
$\textrm{cm}^{-3}~K$ for the plasma pressure. Similarly, assuming that 
the absorbed plasma component is due solely to GH emission, we obtain 
a plasma density ranging from 0.0005 to 0.0014 $cm^{-3}$ and a 
pressure between 3.3$\times10^{3}$ and 5.8$\times10^{3}$ $cm^{-3}K$.

We also used the non-equilibrium plasma model GNEI \citep{Borkowski01}, 
a non-equilibrium model characterized by a constant postsock electron 
temperature and by its ionization age, to fit our data. While we obtained 
a good fit, similar to that shown in Fig.~5, and an electron density 
in the range 
0.013-0.158 $cm^{-3}$, we derived a value for the age of the LB of 
$\leq$~0.9 Myr, which is quite small in comparison with generally accepted 
models (e.g., \citep{Edgar93}).

\section{Comparing {\it Suzaku} and 
{\it XMM-Newton} Observations of the 
Soft X-ray Background.}

The temperature and emission measures we obtained from the {\it Suzaku} data 
are significantly different from those determined from the {\it XMM-Newton} 
analysis in the same pointing directions. For a visual estimate of the 
difference, we folded our {\it Suzaku} model through the {\it XMM-Newton} response 
and compared it with the {\it XMM-Newton} spectra (see Fig.~6). 
The difference in these spectra would be consistent with a time dependent 
component of the foreground emission, attributable to SWCX, which we will 
discuss in detail in the next few sections. 
The excess is clearly significant in both data sets.

So far only a few targets with the proper characteristics for shadow 
experiments have been observed with any of the three major X-ray 
satellites ({\it Chandra}, {\it XMM-Newton}, and {\it Suzaku}). In addition to the 
MBM20 observations discussed, we point out the observations of the 
neutral hydrogen cloud MBM12 performed with {\it Chandra} \citep{Smith05} 
and {\it Suzaku} \citep{Smith06} and that of a relatively dense neutral 
hydrogen filament in the southern galactic hemisphere \citep{Henley08}.

Table~3 summarizes the O~{\tiny VII} and O~{\tiny VIII} flux for all 
the available observations. Data from McCammon et al. 2002 are also 
reported for comparison. In McCammon et al. a high resolution measurement 
over a ~1 sr field of view near the north Galactic pole was performed 
using cryogenic microcalorimeters mounted on a sounding rocket. 
Tables~4 and 5 give a summary dividing the results in foreground and 
background emission. Where a fit with a plasma model has been performed, 
the best fit parameters for temperature and emission measure are also 
reported.

While the amount of available data is limited, we identified a few 
general trends that we want to point out: 
\begin{itemize}
\item Each target has been observed at least twice in the past 8 years, 
but the results from multiple observations of the same target do not 
agree, at various levels, with each other. This is evidence of 
a significant contribution from SWCX, the only component of the diffuse 
X-ray background that should change with time on such a short 
time scale. Moreover, when we separate foreground and background oxygen 
line emission, the component that changes with time seems to be the 
foreground one, while the background does not change, within the errors, 
between different observations of the same target, strengthening the notion
that the variation is due to SWCX. We want to point out, 
however, that multiple observations of the same target have been performed 
with different satellites, i.e., different data reduction analysis, 
background subtraction schemes, etc., which have different systematic 
uncertainties.

\item The change in oxygen line emission between different observations 
of the same target can be used to estimate the typical flux variation 
of the SWCX emission. The O~{\tiny VII} emission varies between 
$1.55\pm0.61$ LU and $4.14\pm0.90$ LU, while the O~{\tiny VIII} 
emission aries between $0.22\pm0.72$ LU and $2.10\pm0.37$ LU. 
The detailed results are reported in Table~6. 

\item High resolution investigations of the diffuse X-ray emission 
have shown that a simple one-temperature plasma in equilibrium cannot 
explain the observed spectra \citep{McCammon02, Sanders01}. However, 
while CCD detectors are a significant step forward from proportional 
counters, their resolution is still quite limited and insufficient 
to investigate the issue. Equilibrium plasma models seem to be still 
sufficient to fit the spectra and, while there have been attempts at 
using more sophisticated models, it is impossible to distinguish 
between them. At this point the available data are adequately fit 
with a plasma thermal emission from the LB, with temperature around 
1 million degrees, and either one or two temperature thermal plasma 
components for the Galactic halo, with temperatures between 2 and 3 
million degrees.

\item Except for the {\it Chandra} observation of MBM12, with its very 
unusual O~{\tiny VIII} emission, all other observations seem to indicate 
that the foreground O~{\tiny VIII} emission is either very small or 
compatible with 0. Typical LB models do not predict  
significant O~{\tiny VIII} emission and this seems to indicate that the 
SWCX component does not normally have any significant emission in 
O~{\tiny VIII} either.

\item When the assumption is made that all the foreground emission is due 
to LB emission, the derived values for the plasma temperature, density, 
and pressure seem to be in good agreement with the predictions from the 
most commonly accepted models of the origin and structure of the LB 
(e.g., Smith and Cox 2001).
\end{itemize}

\section{SWCX Model to Data comparison}

The heliospheric SWCX model we use for our simulations is extensively 
described in Koutroumpa et al. (2006, 2007). This model is a self-consistent 
calculation of the solar wind charge-exchange X-ray line emission for 
any line of sight (LOS) through the heliosphere and for any observation 
date, based on 3-dimensional grids of the inter-stellar (IS) neutral 
species (H and He) distributions in the heliosphere modulated by solar 
activity conditions (gravity, radiation pressure, and ionization processes 
which are anisotropic due to the latitudinal anisotropy of the solar wind 
mass flux and solar radiation). Highly charged heavy solar wind (SW) 
ions are propagated radially through these grids and the charge-transfer 
collision rates are calculated for each of the ion species, including the 
evolution of their density due to charge-transfer with the IS atoms. 
With this process, we establish 3-dimensional emissivity grids for each 
SW ion species, using photon emission yields computed by Kharchenko 
\& Dalgarno (2000) for each spectral line following charge exchange with 
the corresponding neutral species (H and He individually). 
Finally, the X-ray line emission is integrated along any LOS and 
observation geometry (for each observation date) in order to build the 
complete spectrum of SWCX emission in the given direction. 
For comparison to present X-ray observations we use the O~{\tiny VII} 
triplet at 0.57 keV and the O~{\tiny VIII} line at 0.65 keV, as they 
are the strongest spectral features and provide the best signal-to-noise 
ratio for the observations.

We have conducted simulations for each of the MBM20 and Eridanus Hole 
observations accounting as close as possible for average solar activity 
conditions corresponding to the observation period. Solar activity is 
reflected both in the IS neutral distributions (by means of ionization 
rates that are increased and less anisotropic in solar maximum), and in 
the solar wind ionic composition (abundances and charge state distributions) 
and spatial distribution.

Details on the solar activity effect on the neutral H and He distributions, 
along with the latitude-dependant ionization rates used in the model 
for maximum (e.g. 2001), intermediate (e.g. 2003-2004) and minimum 
(e.g. 2007-2008) solar conditions are given in Koutroumpa et al. (2009).

The latitude dependence of the solar wind also affects the highly charged 
heavy ion distribution, where abundances depend on the solar wind type. 
During minimum solar activity, the solar wind is considered to be highly 
anisotropic, with a narrow equatorial zone (within $\pm20^\circ$ of the 
solar equatorial plane) of slow solar wind with an average speed of 
$\sim$400 $km~s^{-1}$ and the fast solar wind emitted from the polar 
coronal holes at a speed of $\sim$700 $km~s^{-1}$. 
The slow solar wind has a proton density of $\sim$6.5~$cm^{3}$ at 1~AU, 
while the fast flow is less dense at $\sim$3.2~$cm^{3}$ at 1~AU. 
At solar maximum, the solar wind spatial distribution is considered to 
be a complex mix of slow and fast wind states that is in general 
approximated with an average slow wind flux. The ionic composition of 
the two flows can be very different with the average oxygen content 
varying from [O/H] = 1/1780 in the slow wind and [O/H] = 1/1550 in the 
fast flow. The charge-state  distributions change as well, with the 
higher charge-states strongly depleted (or even completely absent, as 
for example $O^{+8}$ in the fast solar wind. For our model we adopt 
the oxygen relative abundances published in Schwadron \& Cravens (2000):
($O^{+7}$, $O^{+8}$)=(0.2, 0.07) for the slow wind and 
($O^{+7}$, $O^{+8}$)=(0.07, 0.0) for the fast wind, based on data from 
the Ulysses SWICS instrument.

The {\it XMM-Newton} observations of MBM20 and the Eridanus Hole were 
performed during 2004, which corresponds to intermediate solar 
conditions, while for the {\it Suzaku} observations, performed in 2007-2008, 
solar minimum conditions are most appropriate. The main difference in the 
SW heavy ion distribution between the two periods (two sets of coupled 
observations) is the spatial (latitudinal) distribution of the slow and 
fast solar wind flows. For the {\it Suzaku} simulations (solar minimum) the 
slow SW is expanding in interplanetary space through a $\pm20^\circ$ 
equatorial zone on the solar surface, while the fast SW flow occupies 
the rest of the space. For the intermediate 2003-2004 period 
({\it XMM-Newton} observations) we assume that there is no fast wind flow 
in interplanetary space (same approach as for solar maximum), in order 
to estimate the quiescent (outside potential coronal mass ejection or 
solar flare) upper limit for the resulting SWCX X-ray emission. 
Indeed, as demonstrated in Koutroumpa et al. (2006, 2007), for high 
ecliptic-latitude LOS, as is the case for the MBM20 and Eridanus Hole 
observations (Declination$\sim-38^\circ$), the oxygen line intensity 
decreases from solar maximum to solar minimum conditions as the LOS 
crosses larger fast wind regions where the parent ions are strongly depleted.

In Table~7 we summarize the SWCX model results for the oxygen line 
intensities for the four observations. As expected, model A, which assumes 
average solar wind conditions as described above, predicts a significant 
decrease in the SWCX oxygen line intensities as we progress from near 
solar maximum ({\it XMM-Newton}) to solar minimum ({\it Suzaku}), since we are 
observing at high southern ecliptic latitudes. Also, the model predicts 
a decrease in the heliospheric SWCX emission when shifting the view 
direction from the Eridanus Hole (off-cloud) to the MBM20 (on-cloud) 
direction. This decrease is of the order of 8\% for the {\it XMM-Newton} 
observations and of the order of 30\% up to 45\% (for O~{\tiny VII}) for 
the {\it Suzaku} observations. Such a large difference can be explained by 
the large interval separating the two {\it Suzaku} observations combined 
with the inclination of the equatorial SW zone with respect to the 
ecliptic plane (due to the $7.25^\circ$ inclination of the solar 
axis with respect to the ecliptic axis). Indeed, the {\it Suzaku} 
observations of MBM20 and Eridanus Hole were performed at an observed 
ecliptic longitude of $142^\circ$ and $306^\circ$, respectively, 
separated by six months, while the LOS was pointing at $\sim37^\circ$ 
south. Since the solar equator ascending node is $\Omega=73.67^\circ$, 
the Eridanus Hole LOS was looking through a larger region of the 
oxygen-rich slow solar wind equatorial zone than the MBM20 LOS.

One step to further improve the accuracy of our prediction is to apply 
reasonable assumptions to the SWCX simulations. First, evidence from the 
Ulysses/SWICS $O^{+7}/ O^{+6}$ ratio data (which is a proxy for the 
flow speed/type) during the 2007 (minimum) 
crossing of the equatorial slow wind zone (Fig.~7) shows that this later 
was in fact more extended in latitude (more than $\pm30^\circ$ than 
during the previous solar minimum (1996) that served as a reference 
for the minimum SW conditions applied in the SWCX model. In order to 
investigate the effect of such a possibility we performed a second 
simulation of the {\it Suzaku} Eridanus Hole/MBM20 observations introducing 
a $\pm30^\circ$ slow SW zone as input. The results are also noted in 
Table~7 (Model B). However, in situ measurements with ACE at the L1 
point show unusually low $O^{+7}$ abundances for the {\it Suzaku} observations 
period, almost an order of magnitude lower than the average slow wind 
conditions (11.5\% during the MBM20 observation and 20\% during the 
Eridanus Hole one). For solar maximum ({\it XMM-Newton} observations), the $O^{+7}$ 
abundance in the ACE data does not show significant deviation from average 
slow wind values. $O^{+8}$ measurements are too sparse to allow a 
significant quantitative analysis of the data, and therefore we will 
make no assumption for these data. If we apply 11.5\% and 20\% correction 
factors to the {\it Suzaku}'s MBM20 and Eridanus Hole O~{\tiny VII} line 
intensities predicted from Model B, we obtain the values noted as Model 
B1 in Table~7.

To compare the model results to the observations of MBM20 and the Eridanus 
Hole we must consider that MBM20 absorbs about 75\% of the background 
O~{\tiny VII} emission and about 61\% of the background O~{\tiny VIII} 
emission and therefore we expect a significant contamination from the 
background emission. Table~8 summarizes the final values predicted in the 
SWCX simulations (model B1 from Table~7), along with the measured 
O~{\tiny VII} and O~{\tiny VII} fluxes, the estimated foreground 
(local) flux from Table~4, and the predicted residual cosmic background 
(data minus model).

As the results in Table~8 show, the SWCX O~{\tiny VII} prediction is 
comparable, within one sigma, with the measured local emission. 
Also, the residual O~{\tiny VII} cosmic background has a constant 
value, within error bars, for all on-cloud and off-cloud observations 
that is consistent with the extrapolated background emission 
reported in Table~5. Both results seem to indicate that the 
O~{\tiny VII} foreground emission is dominated by SWCX. This 
conclusion is also supported by a previous application of the 
model to the MBM12 observations \citep{Koutroumpa07}. Due to the 
significantly higher absorption of MBM12, the model results were 
compared directly to the total measured flux and the agreement 
was within 30\%. We point out that this conclusion does not preclude 
the existence or a Local Hot Bubble which is expected to emit 
X-rays primarily at lower energy, in the $1/4$ keV band. 
Models predict a LB O~{\tiny VII} surface brightness of about 
0.25 LU.

The O~{\tiny VIII} results are not as clear, as the measured data 
are consistent with a zero local emission, while the model predicts 
a small, but non-zero emission. However, as mentioned before, the 
ACE $O^{8+}$ data are too sparse and could not be used as input for 
our model. The negligible O~{\tiny VIII} flux could therefore 
simply be caused by a smaller than expected $O^{8+}$ density in the 
solar wind.

\section{Conclusion} 

We used {\it Suzaku} observations of the molecular cloud MBM20 and a low 
neutral hydrogen column density region nearby to separate and 
characterize the foreground and background diffuse X-ray emission. 
We measured a foreground flux of $0.99\pm0.91$ LU and $0.01\pm1.01$ LU 
for O~{\tiny VII} and O~{\tiny VIII} respectively and a background 
flux of $5.10\pm1.79$ LU and $1.42\pm1.74$ LU of O~{\tiny VII} 
and O~{\tiny VIII} respectively.

The comparison with a previous observation of the same regions with 
{\it XMM-Newton} indicates a significant change in the foreground flux 
which we attribute to Solar Wind Charge eXchange. By combining our 
results with similar multiple shadow investigation of the same 
target we find that the O~{\tiny VII} emission varies between 
$1.55\pm0.61$ LU and $4.14\pm0.90$ LU between multiple observations 
of the same target. The O~{\tiny VIII} emission, except for a 
single case with a change of $2.10\pm0.37$ LU, is generally 
compatible with zero, possibly indicating a very low density 
of $O^{8+}$ in the solar wind. 

We also compared our results with a SWCX model to constrain 
its O~{\tiny VII} and O~{\tiny VIII} emission. The model is 
in good agreement with the measured O~{\tiny VII} flux and 
seems to indicate that most of the O~{\tiny VII} foreground 
emission is due to SWCX. This is not necessarily inconsistent 
with the existence of a local hot bubble which is expected to 
emit predominantly at lower energy, in the $1/4$ keV band.

With the limited energy resolution 
of the CCD detectors, the foreground emission can also be modeled with 
an unabsorbed plasma model and the background one with a one 
temperature absorbed plasma model plus an absorbed power law,
as done in previous papers. 
A global fit using both datasets and {\it ROSAT} All Sky Survey Data 
for the same targets is consistent with a foreground plasma 
emission with $T=0.7\times10^{6}$~K and $EM=0.096$~cm$^{-6}$~pc 
and a background plasma emission with $T=2.15\times10^{6}$~K 
and $EM=0.0031$~cm$^{-6}$~pc . We also obtained a good fit by 
using a non-equilibrium plasma model for the foreground emission, 
however the inferred age of the plasma is $\leq0.9$ Myr, 
inconsistent with any Local Bubble model.

\clearpage

\begin{figure}
\begin{center}
\includegraphics[height=8cm]{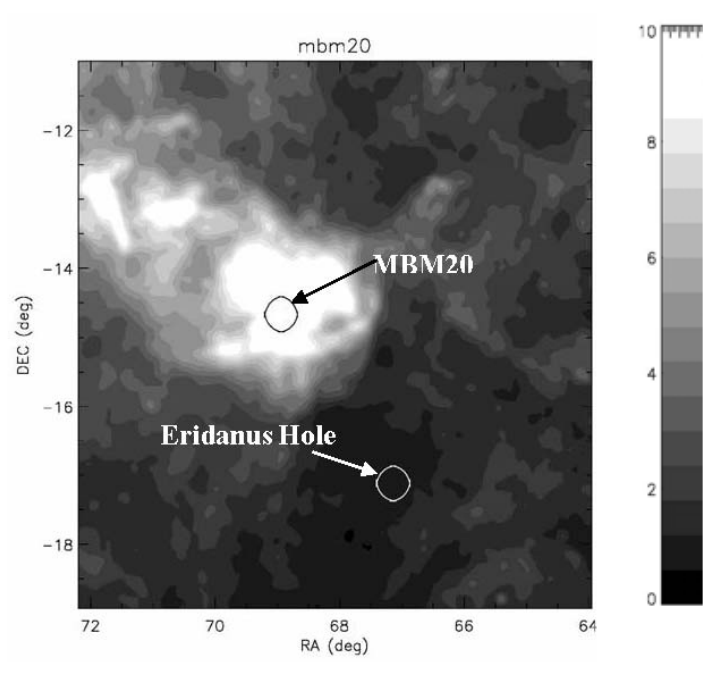}
\end{center}
\caption{IRAS 100$\mu$m map of MBM20 and surroundings showing the two 
pointing used in this investigation.}
\label{fig1}
\end{figure}

\clearpage

\clearpage

\begin{figure}
\begin{center}
\includegraphics[height=8cm]{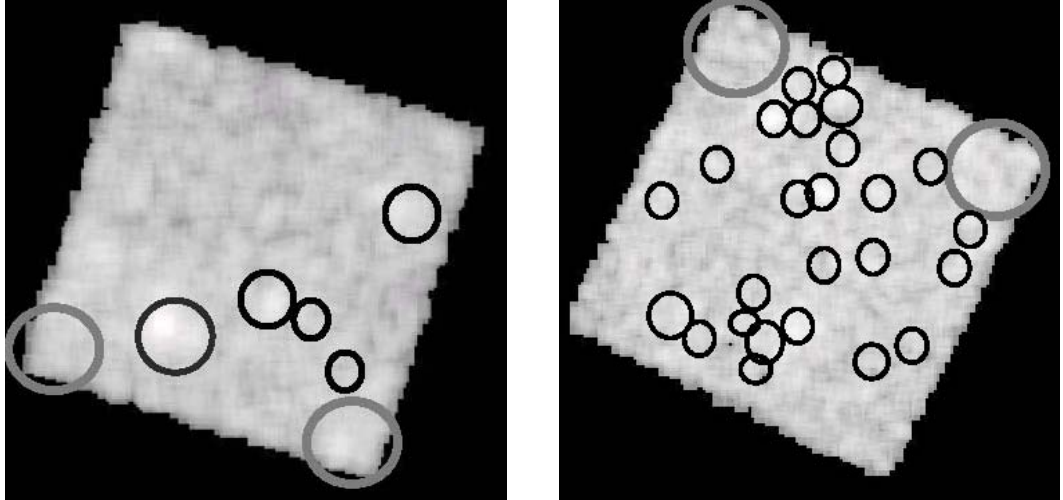}
\end{center}
\caption{XIS1 image of MBM20 ({\it left} and Eridanus Hole ({\it right} in 
the energy range 0.5-2.0 keV. Point sources (black circle) and corners 
of the detector (grey circles) have been removed for the analysis.}
\label{fig2}
\end{figure}

\clearpage

\clearpage

\begin{figure}
\begin{center}
\includegraphics[height=16cm]{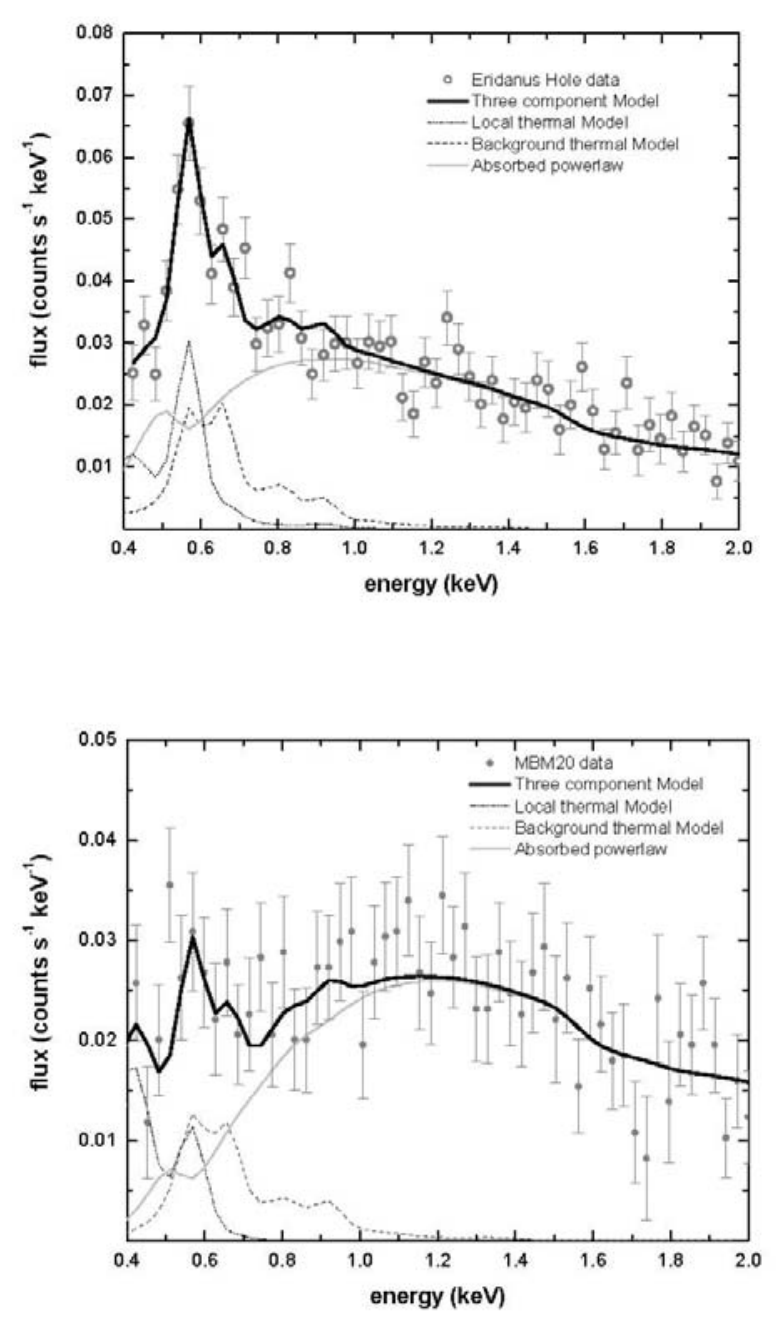}
\end{center}
\caption{Eridanus Hole({\it Top}) and MBM20({\it Bottom}) {\it Suzaku} spectra, 
with the best fitting three component model.}
\label{fig3}
\end{figure}

\clearpage

\clearpage

\begin{figure}
\begin{center}
\includegraphics[height=8cm]{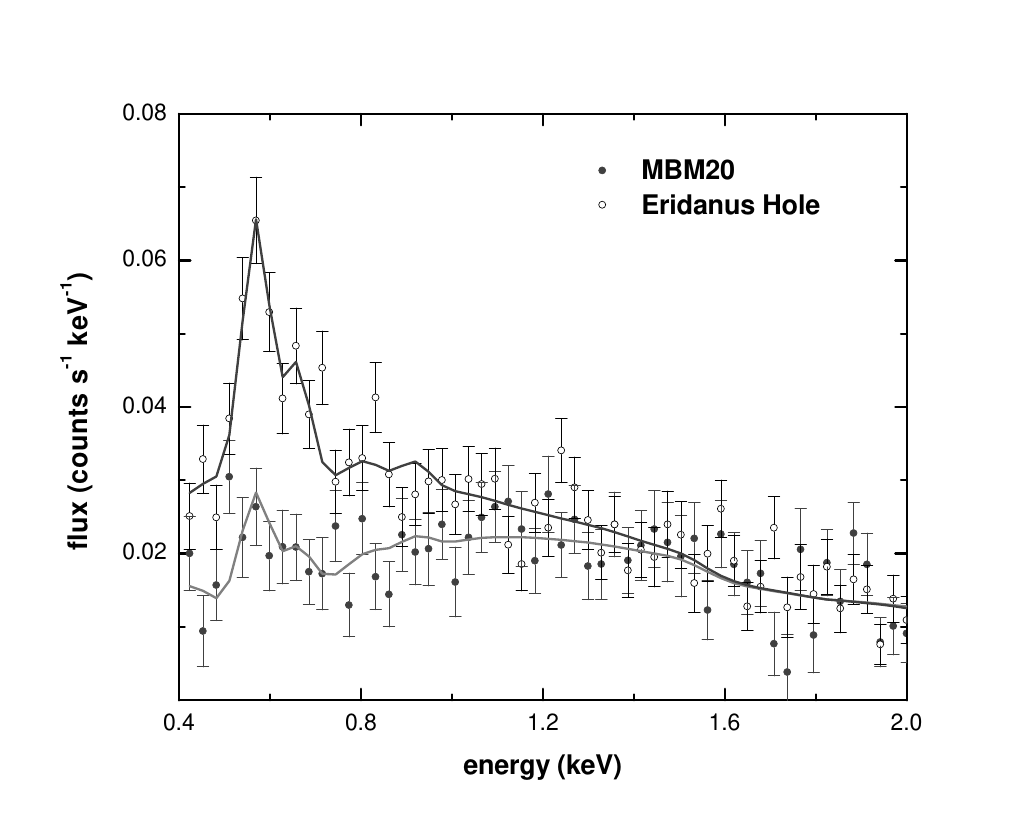}
\end{center}
\caption{Simultaneous fit for MBM20 and Eridanus Hole {\it Suzaku} data.}
\label{fig4}
\end{figure}

\clearpage

\clearpage

\begin{figure}
\begin{center}
\includegraphics[height=8cm]{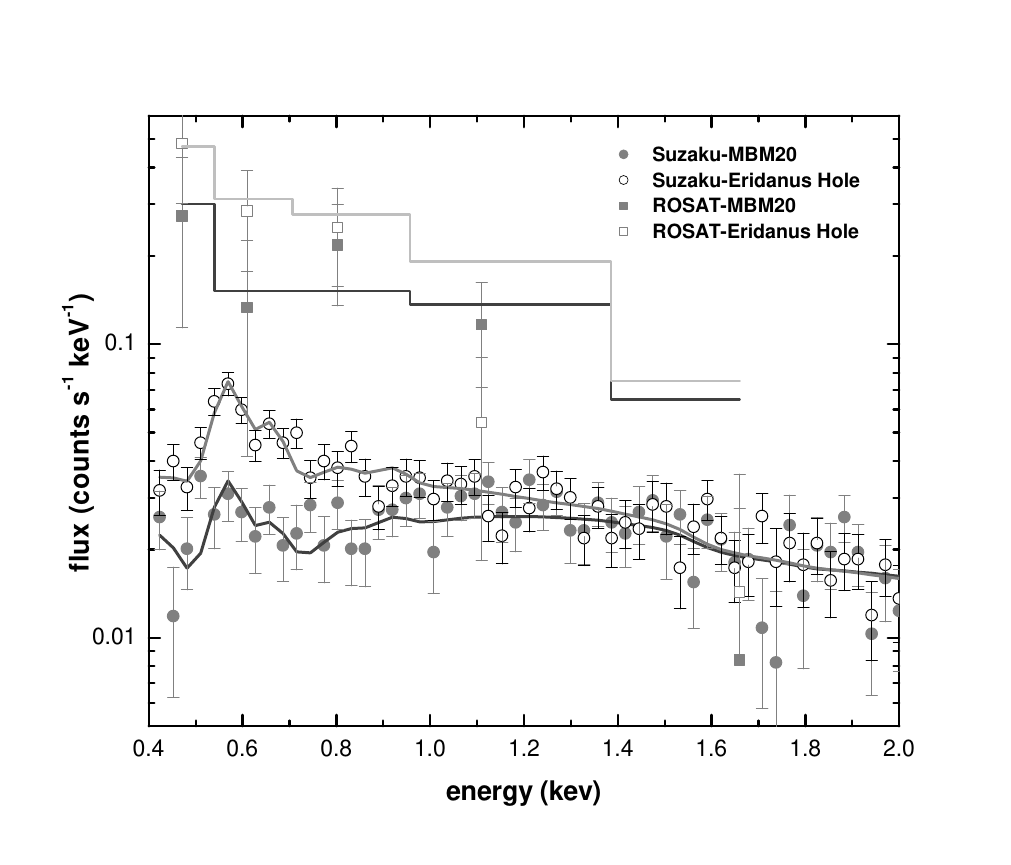}
\end{center}
\caption{Global fits for MBM20 (dark grey) and Eridanus Hole (grey) 
using data from our {\it Suzaku} observations (circles) and RASS (squares).}
\label{fig5}
\end{figure}

\clearpage

\clearpage

\begin{figure}
\begin{center}
\includegraphics[height=16cm]{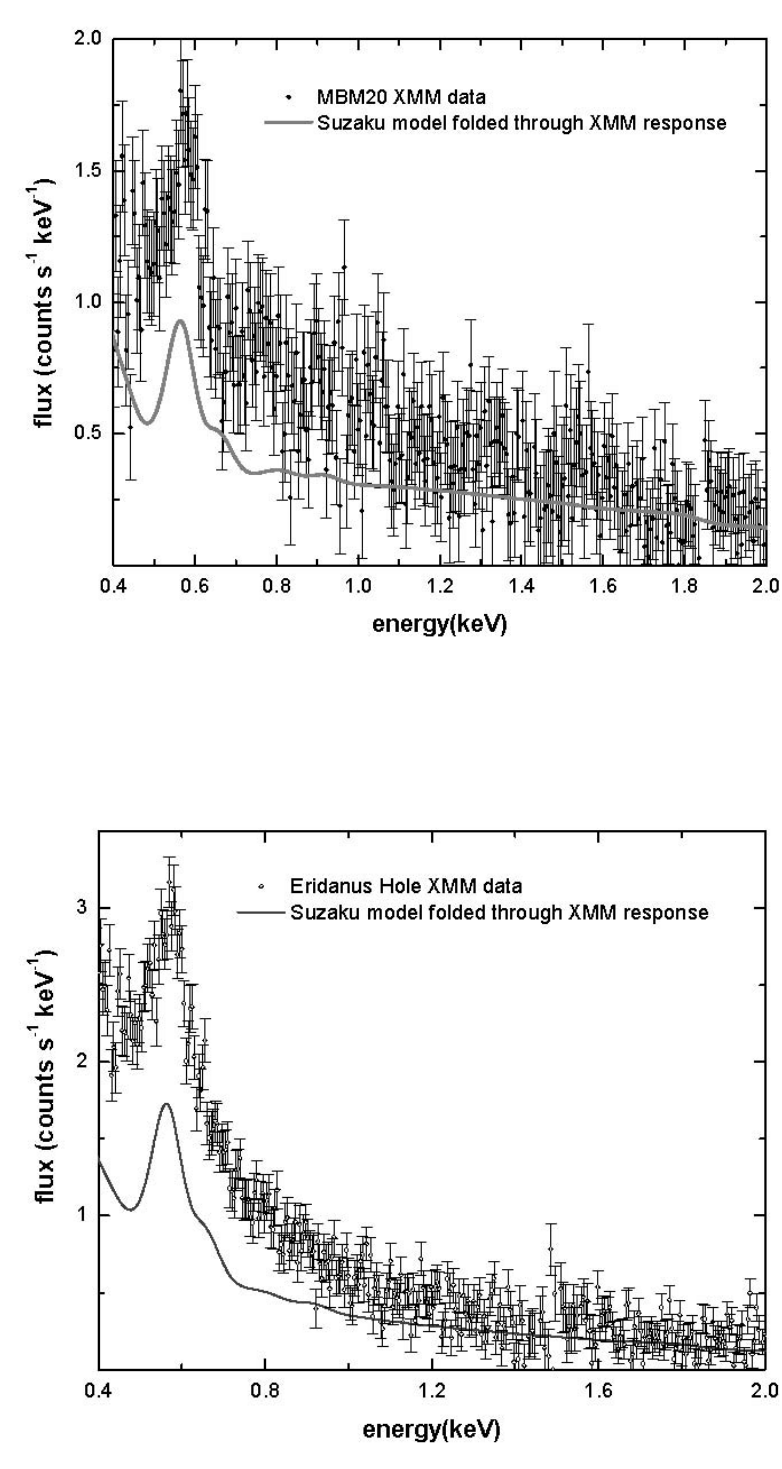}
\end{center}
\caption{{\it Top}: Comparison between {\it XMM-Newton} MBM20 spectra and 
{\it Suzaku} model folded through {\it XMM-Newton} response. 
{\it Bottom}: Same as top for Eridanus Hole.}
\label{fig7}
\end{figure}

\clearpage

\clearpage

\begin{figure}
\begin{center}
\includegraphics[height=6cm]{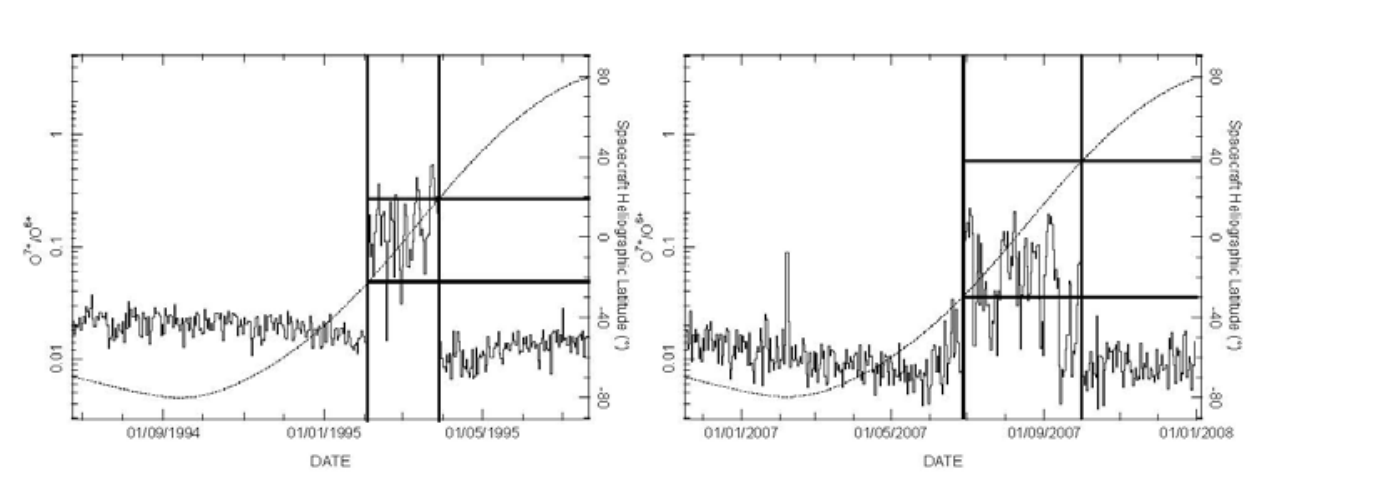}
\end{center}
\caption{{\it Left}: Ulysses/SWICS $O^{+7}/ O^{+6}$ ratio data (plain) 
during the 1995 crossing of the solar equatorial plane. The dotted curve 
marks the spacecraft heliographic latitude on the right axis. 
The vertical and horizontal plain lines denote the limits of the slow solar 
wind equatorial zone in terms of crossing time and heliographic latitude 
respectively. {\it Right}: Same as left, except for the 2007 crossing 
of the solar equatorial plane.}
\label{fig8}
\end{figure}

\clearpage

\begin{deluxetable}{lcccc}
\tablewidth{0pc}
\tablecaption{Details of our {\it Suzaku} observations.}
\tablehead{
\colhead{Target}           & \colhead{Observation ID}      &
\colhead{Start Time(UT)}          & \colhead{End Time(UT)} & \colhead{Exposure(ks)}}
\startdata
MBM20 & 502075010 & 2008-02-11 14:41:19 & 2008-02-14 16:45:11 & 69.2\\
Eridanus Hole & 502076010 & 2007-07-30 00:51:47 & 2007-08-01 05:11:19 & 84.6\\
\enddata
\end{deluxetable}

\clearpage

\clearpage
\begin{deluxetable}{lccccccccc}
\tablecolumns{10}
\tablewidth{0pc}
\tablecaption{Model parameters of the spectral fits}
\tablehead{
\colhead{Dataset(s)}  & \multicolumn{2}{c}{Local component} & \colhead{} & \multicolumn{2}{c}{Galactic Halo} & \colhead{} & \multicolumn{2}{c}{Power Law}  & \colhead{\it {$\chi^2/dof$}}\\
\cline{2-3} \cline{5-6} \cline{8-9}\\
\colhead{} & \colhead{T} & \colhead{E.M.\tablenotemark{a}} & \colhead{} & \colhead{T} & \colhead{E.M.} & \colhead{} & \colhead{$\Gamma$\tablenotemark{b}} & \colhead{Norm\tablenotemark{c}} & \colhead{}\\
\colhead{} & \colhead{($10^{6}~K)/keV$} & \colhead{$cm^{-6}~pc$} & \colhead{} & \colhead{$(10^{6}~K)/keV$} & \colhead{$cm^{-6}~pc$} & \colhead{} & \colhead{} & \colhead{} & \colhead{}}
\startdata
\textbf{MBM20} & 0.78/0.067 & 0.041 & & 1.87/0.16 & 0.0034 & & 1.48 & 8.4 & 167.6/151\\
\textbf{EH} & 1.24/0.106 & 0.007 & & 2.46/0.21 & 0.0016 & & 1.57 & 8.1  & 134.3/151\\
\textbf{(MBM20+EH)\tablenotemark{1}} & 0.83/0.071 & 0.027 & & 2.11/0.18 & 0.0027 & & 1.53 & 8.3  & 304.9/308\\
\textbf{(MBM20+EH)\tablenotemark{2}} & 0.76/0.067 & 0.056 & & 2.12/0.19 & 0.0031 & & 1.33 & 9.1  & 306.2/352\\
\textbf{(MBM20+EH)\tablenotemark{3}} & 1.12/0.096 & 0.0078 & & 2.23/0.19 & 0.0035 & & 2.2 & 14.8  & 768/640\\
\enddata

\tablenotetext{a}{Emission Measure}
\tablenotetext{b}{Index of absorbed power law fit}
\tablenotetext{c}{Normalization of power law fit at 1~keV in units of 
$\textrm{photons~keV}^{-1}~\textrm{s}^{-1}~\textrm{cm}^{-2}~\textrm{sr}^{-1}$ }
\tablenotetext{1}{Suzaku data only}
\tablenotetext{2}{Suzaku and RASS data}
\tablenotetext{3}{XMM-Newton result from Galeazzi et al. 2007}
\end{deluxetable}

\clearpage

\begin{deluxetable}{lccc}
\tablewidth{0pc}
\tablecaption{Summary of the oxygen line emission for MBM12, 
MBM20, and the filament in the southern galactic hemisphere (SGF). 
The data from McCammon et al. 2002 are also reported.}
\tablehead{
\colhead{Experiment}     & \colhead{NH($10^{20}$ $cm^{-2}$)}    &
\colhead{O~{\tiny VII}}  &  \colhead{O~{\tiny VIII}}}
\startdata
\textbf{MBM20} & 15.9 &  & \\
XMM-Newton & & $3.89\pm0.56$ & $0.68\pm0.24$\\
Suzaku & & $2.26\pm0.60$ & $0.56\pm0.48$\\
\textbf{Eridanus Hole} & 0.0086 & & \\
XMM-Newton & & $7.26\pm0.34$ & $1.63\pm0.17$\\
Suzaku & & $5.68\pm1.04$ & $1.32\pm0.79$\\
\textbf{MBM12-on cloud} & 40 & & \\
Chandra & & $1.79\pm0.55$ & $2.34\pm0.36$\\
Suzaku & & $3.34\pm0.26$ & $0.24\pm0.10$\\
\textbf{MBM12-off cloud} & 8.7 & & \\
Suzaku & & $5.68\pm0.42$ & $0.01\pm0.19$\\
\textbf{Henley et al. on filament} & 9.6 & &  \\
XMM-Newton & & $10.65\pm0.80$ & $3.91\pm0.26$\\
Suzaku & & $6.51\pm0.41$ & $2.54\pm0.26$\\
\textbf{Henley et al. off filament} & 1.9 & &  \\
XMM-Newton & & $13.86\pm1.44$ & $2.81\pm0.59$\\
Suzaku & & $10.53\pm0.61$ & $3.21\pm0.31$\\
\textbf{McCammon et al. 2002} & 1.8 & & \\
XQC &  & $4.8\pm0.8$ & $1.6\pm0.4$\\
\enddata
\end{deluxetable}

\clearpage

\begin{deluxetable}{lcccc}
\tablewidth{0pc}
\tablecaption{Summary of the foreground emission for the targets 
discussed in this section.}
\tablehead{
\colhead{Experiment} & \colhead{T} & \colhead{EM} & \colhead{O~{\tiny VII}} & \colhead{O~{\tiny VIII}}\\
\colhead{} & \colhead{$10^{6} K$} & \colhead{$\textrm{cm}^{-6}~pc$} & \colhead{LU} & \colhead{LU}}
\startdata
\textbf{MBM20} &  &  & & \\
XMM & 1.12 & 0.0088 & $2.63\pm0.78$ & $0.03\pm0.43$\\
Suzaku & 0.70 & 0.097 & $0.99\pm0.91$ & $0.01\pm1.01$\\
\textbf{MBM12} &  &  & & \\
Suzaku & $\sim1.2$ &  & $3.34\pm0.26$ & $0.24\pm0.1$\\
Chandra &  &  & $1.79\pm0.55$ & $2.34\pm0.36$\\
\textbf{SGF} &  &  & & \\
Suzaku & 0.95 & 0.0064 & $1.1\pm1.1$ & $1.0\pm1.1$\\
XMM & 1.15 & 0.018 & $6.2\pm2.8$ & $\leq~1$\\
\enddata
\end{deluxetable}

\clearpage

\begin{deluxetable}{lcccc}
\tablewidth{0pc}
\tablecaption{Summary of the background emission for the targets 
discussed in this section.}
\tablehead{
\colhead{Experiment} & \colhead{T} & \colhead{EM} & \colhead{O~{\tiny VII}} & \colhead{O~{\tiny VIII}}\\
\colhead{} & \colhead{$10^{6} K$} & \colhead{$\textrm{cm}^{-6}~pc$} & \colhead{LU} & \colhead{LU}}
\startdata
\textbf{MBM20} &  &  & & \\
XMM & 2.23  & 0.0034 & $5.03\pm0.98$ & $1.68\pm0.53$\\
Suzaku & 2.15  & 0.0031 & $5.10\pm1.79$ & $1.42\pm1.74$\\
\textbf{MBM12} &  &  & & \\
Suzaku &  &  &  $2.34\pm0.33$ & $0.77\pm0.16$\\
\textbf{SGF} &  &  & & \\
Suzaku  & 1.29/3.16 & 0.034/0.0065 & $8.8\pm4.9$ & $2.4\pm1.5$\\
XMM & 0.85/2.69 & 0.17/0.011 & $10.9\pm2$ & \\
\enddata
\end{deluxetable}

\clearpage

\begin{deluxetable}{lcc}
\tablewidth{0pc}
\tablecaption{O~{\tiny VII} and O~{\tiny VIII} variations between 
multiple observations of the same object.}
\tablehead{
\colhead{Target} & \colhead{$\Delta$[O~{\tiny VII}] (LU)} & \colhead{$\Delta$[O~{\tiny VIII}] (LU)}}
\startdata
MBM20 & $1.63\pm0.82$ & $0.12\pm0.54$\\
Eridanus Hole & $1.58\pm1.09$ & $0.31\pm1.33$\\
MBM12 & $1.55\pm0.61$ & $2.10\pm0.37$\\
SGF On-filament & $4.14\pm0.90$ & $1.37\pm0.37$\\
SGF Off-filament & $3.33\pm1.56$ & $0.40\pm0.67$\\
\enddata
\end{deluxetable}

\clearpage

\begin{deluxetable}{lccccccccc}
\tablecolumns{10}
\tablewidth{0pc}
\tablecaption{Model SWCX oxygen line intensites in LU.}
\tablehead{
\colhead{ObsId}     & \colhead{Target} & \multicolumn{2}{c}{Model A} & \colhead{} & \multicolumn{2}{c}{Model B\tablenotemark{a}} & \colhead{} & \multicolumn{2}{c}{Model B1\tablenotemark{b}}\\
\cline{3-4} \cline{6-7} \cline{9-10}\\
\colhead{} & \colhead{} & \colhead{O~{\tiny VII}} & \colhead{O~{\tiny VIII}} & \colhead{} & \colhead{O~{\tiny VII}} & \colhead{O~{\tiny VIII}} & \colhead{} & \colhead{O~{\tiny VII}}& \colhead{O~{\tiny VIII}}}
\startdata
0203900101 & EH & 2.04 & 0.80 &  & 2.04 & 0.80 & & 2.04 &0.80\\
0203900201 & MBM20 & 1.88 & 0.74 &  & 1.88 & 0.74 &  & 1.88 & 0.74\\
502076010 & EH & 1.14 & 0.28 &  & 1.45 & 0.42 &  & 0.29 & 0.42\\
502075010 & MBM20 & 0.81 & 0.16&   & 1.29 & 0.38 &  & 0.15 & 0.38\\
\enddata

\tablenotetext{a}{A larger latitudinal extent ($\pm30^{\circ}$) of slow 
wind heavy ion abundances is assumed for solar minimum (Suzaku). 
{\it XMM-Newton} simulation assumptions remain unchanged.}

\tablenotetext{b}{A real-time $O^{+7}$ measured density is applied to 
model B simulation for the {\it Suzaku} observations. $O^{+7}$ data 
taken in situ at the L1 point are extrapolated to the whole LOS. 
{\it XMM-Newton} O~{\tiny VII} simulations and O~{\tiny VIII} simulations 
remain unchanged.}
\end{deluxetable}

\clearpage

\begin{deluxetable}{lccccccccc}
\tablecolumns{10}
\tablewidth{0pc}
\tablecaption{Data and SWCX model oxygen line intensities in LU. The 
foreground values are from Table 4 for the data and are the 
average of the Eridanus Hole and MBM20 values for the models.}
\tablehead{
\colhead{}     & \colhead{} & \multicolumn{2}{c}{Data} & \colhead{} & \multicolumn{2}{c}{Model B1} & \colhead{} & \multicolumn{2}{c}{Residual}\\
\cline{3-4} \cline{6-7} \cline{9-10}\\
\colhead{Mission} & \colhead{Target} & \colhead{O~{\tiny VII}} & \colhead{O~{\tiny VIII}} & \colhead{} & \colhead{O~{\tiny VII}} & \colhead{O~{\tiny VIII}} & \colhead{} & \colhead{O~{\tiny VII}}& \colhead{O~{\tiny VIII}}}
\startdata
 & EH & $7.37\pm0.34$ &  $1.73\pm0.17$ & & 2.04 & 0.80 & & $5.23\pm0.34$ &  $0.83\pm0.17$ \\
XMM & MBM20 & $3.59\pm0.56$ &  $0.72\pm0.24$ & &1.88 & 0.74 & & $2.00\pm0.56$ & $\sim0$ \\
& Foreground & $2.63\pm0.78$ &  $0.03\pm0.43$ & & 1.96 & 0.77 & & $0.67\pm0.78$ & $\sim0$ \\
\\
& EH & $6.68\pm1.04$ &  $1.32\pm0.79$ & & 0.29 & 0.42 & & $5.39\pm1.04$ &  $0.9\pm0.79$ \\
Suzaku & MBM20 & $2.60\pm0.60$ &  $0.57\pm0.48$ & & 0.15 & 0.38 & & $2.11\pm0.6$ & $0.18\pm0.48$ \\
& Foreground & $0.99\pm0.91$ &  $0.01\pm1.01$ & & 0.22 & 0.40 & & $0.77\pm 0.91$ & $\sim 0$ \\ 
\enddata
\end{deluxetable}

\end{document}